\documentclass[manuscript]{aastex}

\usepackage{graphicx}
\usepackage{epstopdf}

\shorttitle{Magnesium in Europa's Atmosphere}
\shortauthors{H\"orst and Brown}

\begin{document}

%% LaTeX will automatically break titles if they run longer than
%% one line. However, you may use \\ to force a line break if
%% you desire.

\title{A Search for Magnesium in Europa's Atmosphere}

%% Use \author, \affil, and the \and command to format
%% author and affiliation information.
%% Note that \email has replaced the old \authoremail command
%% from AASTeX v4.0. You can use \email to mark an email address
%% anywhere in the paper, not just in the front matter.
%% As in the title, use \\ to force line breaks.
\author{S.M. H\"orst\altaffilmark{1}, 
M.E. Brown\altaffilmark{2}}
\altaffiltext{1}{Cooperative Institute for Research in Environmental Sciences, University of Colorado-Boulder, Boulder, CO, USA}
\altaffiltext{2}{Division of Geological and Planetary Sciences, California Institute of Technology, Pasadena, CA, USA}
\email{sarah.horst@colorado.edu}

%% Notice that each of these authors has alternate affiliations, which
%% are identified by the \altaffilmark after each name.  Specify alternate
%% affiliation information with \altaffiltext, with one command per each
%% affiliation.

%\altaffiltext{1}{Joint Appointment, Lunar and Planetary Laboratory, The University of Arizona,1629 E. University Blvd., Tucson, AZ 85721, USA.}

%% Mark off your abstract in the ``abstract'' environment. In the manuscript
%% style, abstract will output a Received/Accepted line after the
%% title and affiliation information. No date will appear since the author
%% does not have this information. The dates will be filled in by the
%% editorial office after submission.

\begin{abstract}
Europa's tenuous atmosphere results from sputtering of the surface. The trace element composition of its atmosphere is therefore related to the composition of Europa's surface. Magnesium salts are often invoked to explain Galileo Near Infrared Mapping Spectrometer spectra of Europa's surface, thus magnesium may be present in Europa's atmosphere. We have searched for magnesium emission in Hubble Space Telescope Faint Object Spectrograph archival spectra of Europa's atmosphere. Magnesium was not detected and we calculate an upper limit on the magnesium column abundance. This upper limit indicates that either Europa's surface is depleted in magnesium relative to sodium and potassium, or magnesium is not sputtered as efficiently resulting in a relative depletion in its atmosphere. 
\end{abstract}

%% Keywords should appear after the \end{abstract} command. The uncommented
%% example has been keyed in ApJ style. See the instructions to authors
%% for the journal to which you are submitting your paper to determine
%% what keyword punctuation is appropriate.


%% From the front matter, we move on to the body of the paper.
%% In the first two sections, notice the use of the natbib \citep
%% and \citet commands to identify citations.  The citations are
%% tied to the reference list via symbolic KEYs. The KEY corresponds
%% to the KEY in the \bibitem in the reference list below. We have
%% chosen the first three characters of the first author's name plus
%% the last two numeral of the year of publication as our KEY for
%% each reference.

%% Authors who wish to have the most important objects in their paper
%% linked in the electronic edition to a data center may do so by tagging
%% their objects with \objectname{} or \object{}.  Each macro takes the
%% object name as its required argument. The optional, square-bracket 
%% argument should be used in cases where the data center identification
%% differs from what is to be printed in the paper.  The text appearing 
%% in curly braces is what will appear in print in the published paper. 
%% If the object name is recognized by the data centers, it will be linked
%% in the electronic edition to the object data available at the data centers  
%%
%% Note that for sources with brackets in their names, e.g. [WEG2004] 14h-090,
%% the brackets must be escaped with backslashes when used in the first
%% square-bracket argument, for instance, \object[\[WEG2004\] 14h-090]{90}).
%%  Otherwise, LaTeX will issue an error. 

\section{Introduction}

Jupiter's moon Europa likely possesses a global subsurface salty ocean \citep{Anderson:1998, Carr:1998, Kivelson:2000}. Although we cannot currently directly measure the ocean composition, if the ocean is in communication with the surface then evaporites may be present on the surface where their composition is accessible to remote sensing. Knowledge of the surface composition therefore plays an important role in ascertaining whether, and to what extent, the subsurface ocean communicates with the surface, and, if the ocean is in communication with the surface, then the surface composition should also provide information about composition of the internal ocean. 

	The Near Infrared Mapping Spectrometer (NIMS), carried by the Galileo spacecraft \citep{Carlson:1992}, obtained spectra of Europa's surface from 0.7 to 5.2 $\mu$m and found evidence for the spectral signatures of abundant non-water ice material. These spectral signatures were initially interpreted as indicative of the presence of hydrated salts, such as hexahydrite (MgSO$_{4}\bullet$6H$_{2}$O) and epsomite (MgSO$_{4}\bullet$7H$_{2}$O), assumed to originate from a subsurface ocean \citep{McCord:1998, McCord:2001}, but it was later shown that hydrated sulfuric acid (H$_{2}$SO$_{4}\bullet$n-H$_{2}$O), an expected product of Europa's irradiation, could also account for the spectral signatures \citep{Carlson:1999, Carlson:2002, Strazzulla:2007}. The presence of other sulfur species (SO$_{2}$ and sulfur allotropes) detected on the surface, primarily on the heavily bombarded trailing hemisphere \citep{Lane:1981, Carlson:2009, Paranicas:2001, Paranicas:2002} indicates that chemistry driven by sulfur ions from Io affects the chemical composition of the surface. However, unambiguous interpretation of the NIMS spectra is not possible due to their relatively low resolution, which precludes the presence of distinct spectral features that would uniquely identify surface constituents \citep{Carlson:2009}. The relative contribution, if any, of evaporites from the subsurface ocean and the composition of those evaporites is still unknown. 
		
	The atmosphere of Europa results primarily from sputtering of the surface by energetic particles from the Jovian magnetosphere. The composition of the atmosphere strongly reflects the composition of the surface; examination of Europa's atmosphere can be used to investigate its surface. Since the surface is dominated by water ice, the principal constituents of Europa's atmosphere are water and its products (see e.g. \citet{McGrath:2009}). The atmosphere also contains traces of other elements; both sodium and potassium have been detected in Europa's atmosphere \citep{Brown:1996, Brown:2001}. While the detection of sodium and potassium could indicate the presence of salts on Europa's surface, it is also possible that they are simply contaminants carried from Io \citep{Brown:1996}. The sodium-to-potassium ratio at Europa is higher than that at Io contrary to the expectation from simple contamination \citep{Brown:2001}. However, as \citet{Carlson:2009} point out that the details of implantation and sputtering are sufficiently uncertain that the ratio difference is not a definitive test.

	In more recent spectral models of NIMS data magnesium sulfates dominate much of the non-water ice component of the surface \citep{Dalton:2005, Dalton:2007, Dalton:2012, Dalton:2012b}. Such an enrichment of magnesium on the surface should result in a corresponding enrichment in the sputtered atmosphere of Europa. Here we use archival spectra from the Hubble Space Telescope Telescope Faint Object Spectrograph (HST-FOS) to search for the signature of the strong 2852 {\AA} resonant scattering emission line of magnesium in Europa's atmosphere. Comparison with known abundances of sodium and potassium in the atmosphere allows us to place constraints on the presence of magnesium on the surface of Europa.	
	
\section{Observations and Data Reduction}

The Hubble Space Telescope observed Europa on 4 July 1994 and 29 July 1994 using the Faint Object Spectrograph. The data include two 35 second spectra centered on Europa, five 924 second spectra offset 8.8 Europa radii (R$_E$) south of Europa and three 1100 second spectra offset 14.4 R$_E$ south of Europa. The spectra were obtained using the blue detector and the A1 aperture, which is $3.66"\times1.29"$ ($1.11\times10^{-10}$ SR). FOS was operated in the image mode using the G270H grating. The spectra cover 2200 to 3200 {\AA}, which includes the magnesium emission line at 2852 {\AA}. Table 1 contains details of the observations. 

Due to problems with the data headers, the archival data are not correctly processed by the Space Telescope Science Institute's automated calibration pipeline POA CALFOS. We instead process
the data manually using the provided post operational archive calibration files. We convert the raw count rate of the spectrograph, in counts s$^{-1}$ diode$^{-1}$, to the standard units of ergs cm$^{-2}$ s$^{-1}$ {\AA}$^{-1}$ by first correcting for diode-to-diode sensitivity variations by multiplying by the flatfield response file and then multiplying by the inverse sensitivity curves. No attempt is made at background subtraction, as we fit and remove scattered light and background in our subsequent analysis. We confirmed our manual reduction routines by performing the same procedures on data that was processed with POA CALFOS.

Figure 1 shows the spectrum centered on Europa and the spectra at the two offset positions. The spectra at both offset positions, while orders of magnitude fainter than that of Europa, are dominated by scattered light from Europa. Visual inspection of the raw FOS spectra does not reveal any emission features from Europa's atmosphere. As can be seen from Figure 1, the scattered light fills the aperture more fully than the 0.87 arcsecond diameter Europa, and spectral lines are thus broader. Depending on the spatial structure of the scattered light, there is also a chance that the scattered light would appear shifted in wavelength compared to the Europa spectrum. Finally, the possibility exists of  stray light scattered in the spectrograph. To account for these factors, we fit each scattered light spectrum with a broadened, shifted, scaled spectrum of Europa with a constant background added. We perform a $\chi^2$ minimization to find the best fit which minimizes the residuals between 2600 and 3000 \AA\ while excluding the region within 20 \AA\ of the expected 2852 \AA\ resonant scattering line. Figure 2 shows the spectra at the offset locations before and after subtraction of scattered light. The subtracted spectra are generally excellent in most regions, but do show some uncorrected systematic variations. 

\section{Data and Results}

Atmospheric emission from Europa will fill the FOS aperture, causing the monochromatic emission line to spread over the full 12 diode width of the A1 aperture, which corresponds to a line width of $\sim$24 \AA. To better see such a broad emission line, we bin the data to 12 \AA\ per pixel. A line filling the aperture will thus have a 2 pixel width in the binned spectrum. Figure 3 shows the binned spectra at the two offset positions. No emission at the 2852 \AA\ location of the resonant scattering line of magnesium can be seen. Because of the uncorrected systematic errors in the subtraction, we cannot compute a statistically rigorous upper limit to the presence of an emission line, but we instead simply add artificial emission at 2852 \AA\ until it reaches a point where we would confidently declare a detection. This modeled lines that would be detectable are shown in Figure 3. Our upper limits to monochromatic emission at 2852 \AA\ in the offset spectra are $2.5 \times 10^{-15}$ erg cm$^{-1}$ s$^{-1}$ at 8.8 R$_E$ and $1.0 \times 10^{-15}$ erg cm$^{-1}$ s$^{-1}$ at 14.4 R$_E$, or an emission strength of 40 R at 8.8 R$_E$ and 16 R at 14.4 R$_E$ (1 Rayleigh (R) = 10$^{6}/4\pi \times$ photons cm$^{-2}$ sec$^{-1}$ SR$^{-1}$). 

We converted the intensity upper limits to upper limits on the magnesium column density using a photon scattering coefficient (\textit{g}) of 1.8 $\times$ 10$^{-3}$ photon s$^{-1}$ atom$^{-1}$. The \textit{g-}value was calculated using the method described in \citet{Chamberlin:1987} using magnesium oscillator strengths from \citet{Haynes:2012} and solar flux from \citet{Hearn:1983} doppler shifted to account for Europa's heliocentric velocity at the time of the observations (see Table 1). These intensity limits imply upper limits for the magnesium column density of $2\times10^{10}$ cm$^{-2}$ at 8.8 R$_E$ and $9\times10^{9}$ cm$^{-2}$ at 14.4 R$_E$. 

\section{Conclusions}

We searched archival HST-FOS spectra of Europa's atmosphere for the 2852 \AA\ magnesium emission feature. We were unable to detect magnesium and, instead, derive upper limits of $2\times10^{10}$ cm$^{-2}$ at 8.8 R$_E$ and $9\times10^{9}$ cm$^{-2}$ at 14.4 R$_E$ on the column abundance of magnesium. Comparison with the sodium and potassium column abundances measured by \citet{Brown:2001} yields upper limits to the magnesium to sodium ratio of $\sim$10 at 8.8 R$_E$ and $\sim$7 at 14.4 R$_E$ and upper limits to the magnesium to potassium ratio of $\sim$280 at 8.8 R$_E$ and $\sim$224 at 14.4 R$_E$ (see Table 2). The meteoritic (CI chondrites) and cosmic Mg/Na are 19 and 18 respectively \citep{Cox:1999, Lodders:1998} thus it appears that Europa's atmospheric Mg/Na is at least a factor of 2 lower than meteoritic and cosmic abundances. The upper limit on Mg/K in Europa's atmosphere is also less than cosmic.

The lower than cosmic abundance of magnesium in Europa's atmosphere has two possible explanations, either magnesium is not present in surface concentrations sufficient to result in a detection in Europa's atmosphere or magnesium is abundant on the surface but is being sputtered into Europa's atmosphere at a lower rate than sodium and potassium. Recent models of spectra of Europa's surface generally include magnesium and sodium sulfate salts (such as hexahydrite (MgSO$_{4}\bullet$6H$_{2}$O), mirabilite Na$_{2}$SO$_{4}\bullet$10H$_{2}$O, and bloedite Na$_{2}$Mg(SO$_{4}$)$_{2}\bullet$4H$_{2}$O) in Mg/Na ratios that are lower than our calculated upper limits for Europa's atmosphere \citep{Dalton:2005, Dalton:2007, Dalton:2012, Dalton:2012b}. Models of Europa's ocean result in high concentrations of both magnesium and sodium with a Mg/Na ratio that is approximately half of our calculated upper limit \citep{Zolotov:2009}.

However, it is also possible that a significant amount of magnesium is present on Europa's surface but it is not being sputtered into the atmosphere. Although magnesium has recently been detected in the atmosphere of Mercury \citep{McClintock:2009}, it has not been detected in the sputtered atmospheres of the Moon \citep{Stern:1997} and Io \citep{Na:1998}. The Mg/Na and Mg/K ratios for the exospheres of Mercury and the Moon are shown in Table 2. Since the abundance of magnesium in the lunar regolith is known, it is possible to model the magnesium abundance in an atmosphere created by sputtering. Based on a stoichiometric atmosphere model \citep{Flynn:1996}, \citet{Stern:1997} find that the magnesium in the lunar atmosphere is depleted by at least factor of 10 based on the known lunar surface composition. The discrepancy between the known abundances of magnesium on the lunar surface and the inability to detect magnesium in the lunar atmosphere has yet to be explained. Although magnesium's presence in the atmosphere of Mercury is not surprising given that magnesium bearing minerals exist on the surface, the relative abundance and distribution of magnesium compared to sodium and calcium indicate that sputtering processes may be different for different minerals or atoms \citep{McClintock:2009, Vervack:2010}. The lunar and Mercurian examples suggest that magnesium may sputter less efficiently than sodium and potassium, a property that has been demonstrated experimentally for some Europa relevant salts; sodium and potassium sulfates are more easily sputtered than magnesium sulfates \citep{McCord:2001}. This difference in sputtering efficiency could result in an enhancement of magnesium salts on the surface relative to potassium and sodium salts and a corresponding depletion in magnesium relative to potassium and sodium in Europa's atmosphere. Although the sodium in Europa's atmosphere is mostly endogenic \citep{Brown:2001}, it is possible that Iogenic sodium implanted on Europa's surface could also affect the ratio. 

The upper limit we have calculated on the Mg/Na ratio in Europa's atmosphere is lower than ratio in CI chondrites by at least a factor of two, which suggests that the trace elements observed in Europa's atmosphere are not meteoritic in origin, assuming similar sputtering efficiencies. However, for the upper limit on the magnesium abundance to provide a stronger constraint on the surface composition, and its origin, more stringent upper limits from future observations are required.

\acknowledgments
This work was initially supported by a Summer Undergraduate Research Fellowship from the California Institute of Technology funded by a trust established by J. Edward Richter. SMH is supported by NSF Astronomy and Astrophysics Postdoctoral Fellowship AST-1102827.

\clearpage

%% Use the figure environment and \plotone or \plottwo to include
%% figures and captions in your electronic submission.
%% To embed the sample graphics in
%% the file, uncomment the \plotone, \plottwo, and
%% \includegraphics commands
%%
%% If you need a layout that cannot be achieved with \plotone or
%% \plottwo, you can invoke the graphicx package directly with the
%% \includegraphics command or use \plotfiddle. For more information,
%% please see the tutorial on "Using Electronic Art with AASTeX" in the
%% documentation section at the AASTeX Web site,
%% http://www.journals.uchicago.edu/AAS/AASTeX.
%%
%% The examples below also include sample markup for submission of
%% supplemental electronic materials. As always, be sure to check
%% the instructions to authors for the journal you are submitting to
%% for specific submissions guidelines as they vary from
%% journal to journal.

%% This example uses \plotone to include an EPS file scaled to
%% 80% of its natural size with \epsscale. Its caption
%% has been written to indicate that additional figure parts will be
%% available in the electronic journal.

\begin{figure}
\plotone{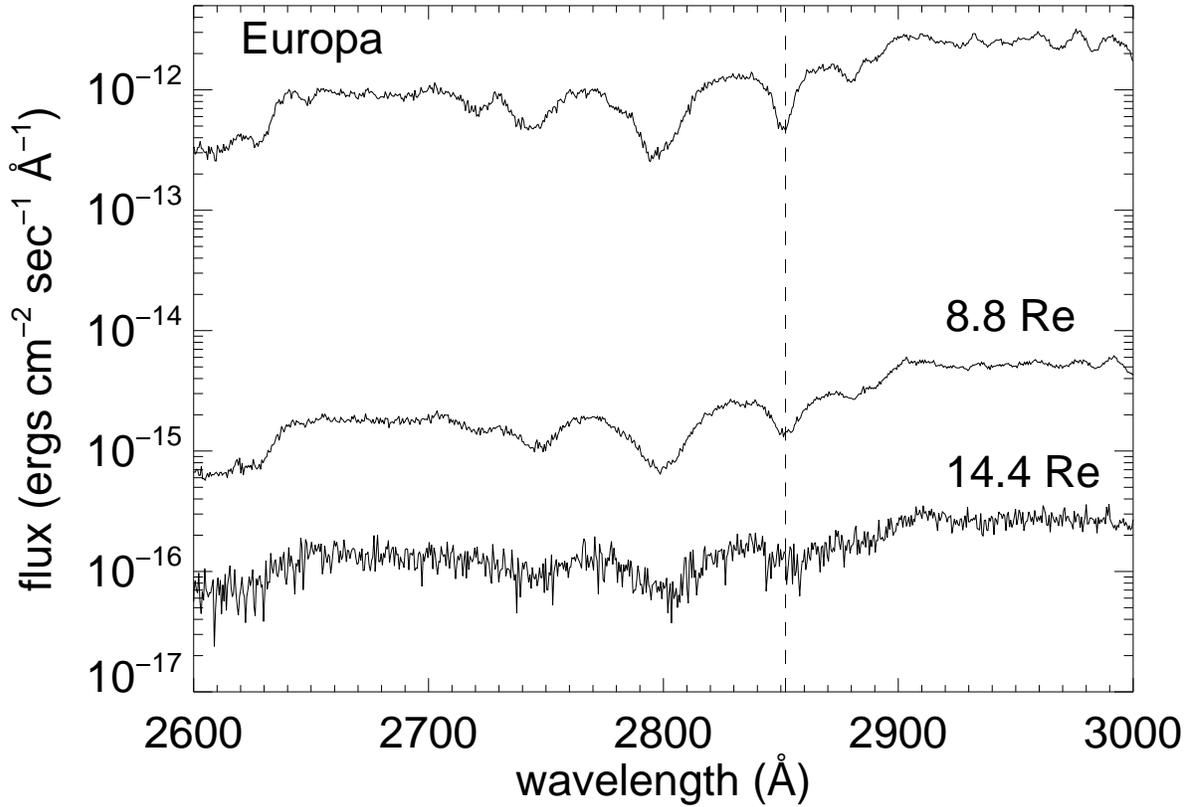}
\caption{HST-FOS spectra of Europa and its atmosphere. The top spectrum was obtained centered on Europa (top), while the middle and bottom spectra were obtained offset by 8.8 R$_{E}$ and 14.4 R$_{E}$ respectively. The dashed line indicates the wavelength of the Mg resonant scattering line at 2852 \AA.}
\end{figure}

\begin{figure}
\plotone{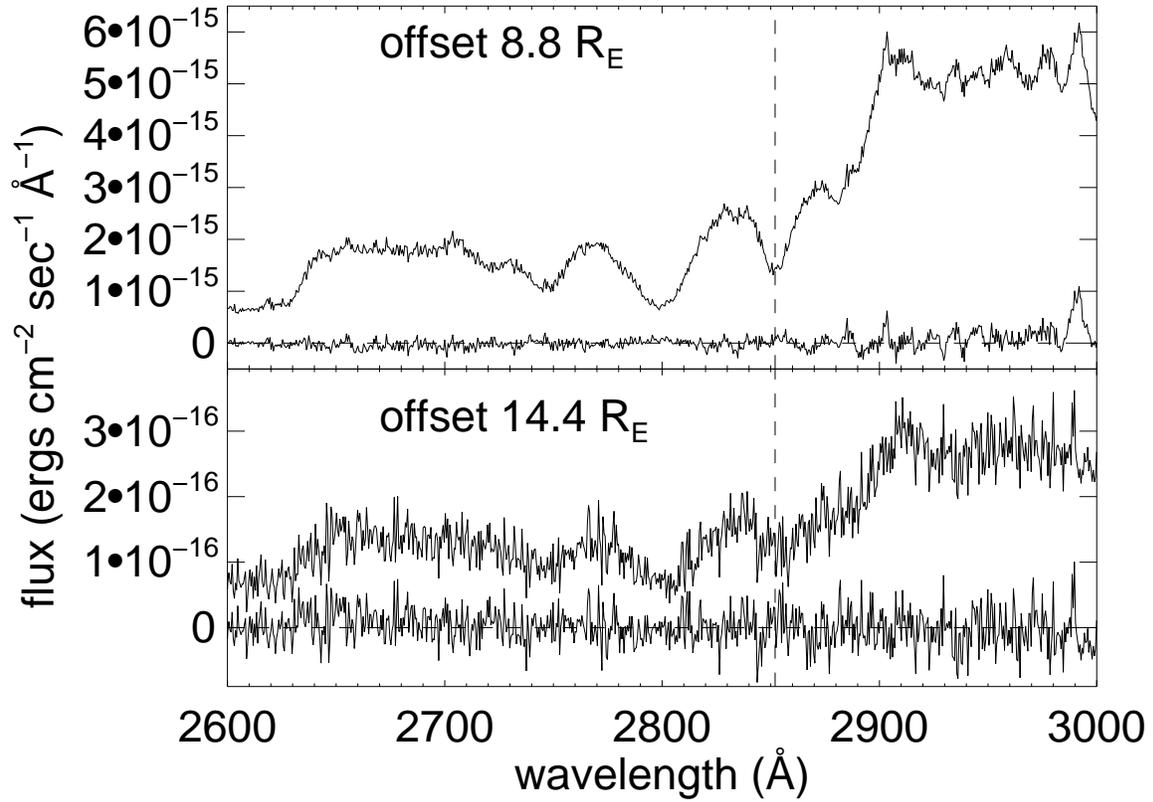}
\caption{Spectra of Europa's extended atmosphere before and after subtraction of scattered light. The dashed line indicates the wavelength of the Mg resonant scattering line at 2852 \AA. }
\end{figure}

\begin{figure}
\plotone{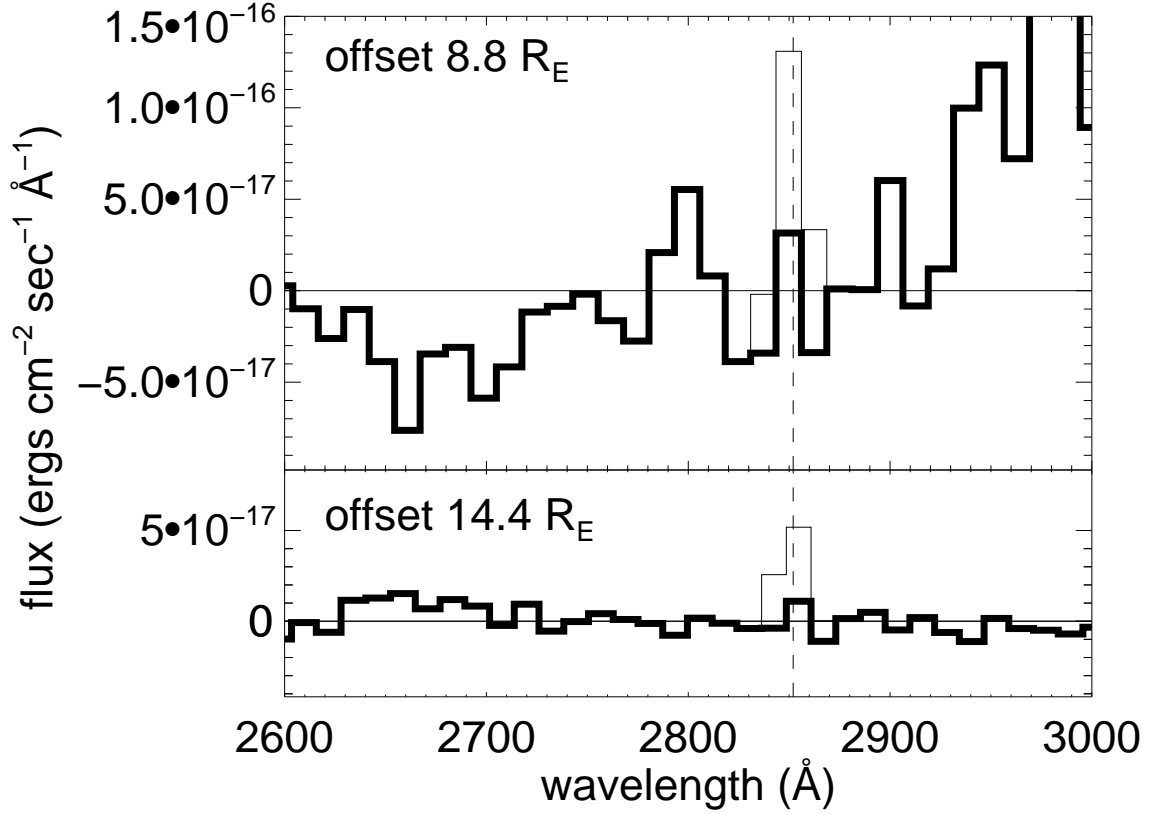}
\caption{Background subtracted spectra binned to the resolution of an emission line filling the aperture (thick lines). The dashed line shows the wavelength of the 2852 \AA\ resonant scattering line of magnesium. The thin lines show the minimum detectable flux and are the basis of our upper limits. }
\end{figure}

\clearpage

\renewcommand{\baselinestretch}{1}

\begin{deluxetable}{rrcrrr}
\tabletypesize{\small} 
\tablenum{1} 
\tablecolumns{6}
\tablewidth{0pt}
\tablecaption{HST-FOS Observations of Europa }\label{Table:observations}
\tablehead{\colhead{Date}&\colhead{Time}&\colhead{Offset from}&\colhead{Exp}&\colhead{$\dot{r}$}&\colhead{Phase}\\
&\colhead{(UTC)}&\colhead{Europa (arcsec)}&\colhead{(s)}&\colhead{(km/s)}&\colhead{($^{\circ}$)}}
\startdata
 4 July 1994 &3:58&	0	&35&	12.62&	263.12\\ 
 4 July 1994 &4:05&	3.85	&924&	12.65&	264.17\\
 4 July 1994 &5:30&	3.85	&924&	13.01&	270.16\\
 4 July 1994 &5:50&	3.85	&924&	13.08&	271.56\\
 4 July 1994 &7:06&	3.85	&924&	13.25&	276.91\\
 4 July 1994 &7:27&	3.85	&924&	13.27&	278.39\\
29 July 1994 &4:17&	5.86	&1100&	13.27&	278.09\\
29 July 1994 &5:36&	5.86	&1100&	13.23&	283.65\\
29 July 1994 &6:00&	5.86	&1100&	13.32&	285.34\\
\enddata
\tablecomments{All observations used the G270H grating, image mode, blue detector, and A1 aperture.}
\end{deluxetable}

\clearpage

\begin{deluxetable}{lrl}
\tabletypesize{\small} 
\tablenum{2} 
\tablecolumns{3}
\tablewidth{0pt}
\tablecaption{Magnesium to Sodium and Magnesium to Potassium Ratios in the Solar System}\label{Table 2}
\tablehead{
\colhead{Object}&\colhead{Mg/Na}&\colhead{Reference}}
\startdata
Offset 8.8 R$_E$ &$<$10&This work, \citet{Brown:2001}\\
Offset 14.4 R$_E$ &$<$7&This work, \citet{Brown:2001}\\
Lunar exosphere & $<$170& \citet{Stern:1997,Stern:1999}, \citet{Potter:1998}\\ %fix this citation
Mercury exosphere&0.05-2.5&\citet{Vervack:2010}\\
Cosmic abundance & 18 & \citet{Cox:1999}\\
CI chondrite&19&\citet{Lodders:1998}\\
Terrestrial oceans & 0.11 & \cite{Haynes:2012}\\
Terrestrial crust & 0.93 & \cite{Haynes:2012}\\[1mm]
\hline
\hline
\noalign{\vskip 2mm} 
\multicolumn{1}{c}{Object}&\multicolumn{1}{c}{Mg/K}&\multicolumn{1}{c}{Reference}\\[1mm]
\hline
\noalign{\vskip 2mm} 
Offset 8.8 R$_E$ & $<$280&This work, \citet{Brown:2001}\\
Offset 14.4 R$_E$ & $<$224 &This work, \citet{Brown:2001}\\
Lunar exosphere & $<$350 & \citet{Stern:1997,Stern:1999}, \citet{Potter:1988}\\ %fix this citation
Mercury exosphere&2-20&\citet{Vervack:2010}, \citet{Potter:2002}\\
Cosmic abundance & 275  & \cite{Cox:1999}\\
CI chondrite&176&\citet{Lodders:1998}\\
Terrestrial oceans & 5.2 & \citet{Haynes:2012}\\
Terrestrial crust & 1.8 & \citet{Haynes:2012}\\
\enddata
\end{deluxetable}

%% If the table is more than one page long, the width of the table can vary
%% from page to page when the default \tablewidth is used, as below.  The
%% individual table widths for each page will be written to the log file; a
%% maximum tablewidth for the table can be computed from these values.
%% The \tablewidth argument can then be reset and the file reprocessed, so
%% that the table is of uniform width throughout. Try getting the widths
%% from the log file and changing the \tablewidth parameter to see how
%% adjusting this value affects table formatting.

%% The \dataset{} macro has also been applied to a few of the objects to
%% show how many observations can be tagged in a table.

%% Tables may also be prepared as separate files. See the accompanying
%% sample file table.tex for an example of an external table file.
%% To include an external file in your main document, use the \input
%% command. Uncomment the line below to include table.tex in this
%% sample file. (Note that you will need to comment out the \documentclass,
%% \begin{document}, and \end{document} commands from table.tex if you want
%% to include it in this document.)

%% \input{table}

%% The following command ends your manuscript. LaTeX will ignore any text
%% that appears after it.

\end{document}